# Exact Network Reconstruction from Consensus Signals and One Eigenvalue

Enzo Fioriti[1], Stefano Chiesa[1], Fabio Fratichini[1]

[1]ENEA, CR Casaccia, UTTEI-ROB, Roma, Italy
vincenzo.fioriti@enea.it

## ABSTRACT

*The basic inverse problem in spectral graph theory consists in determining the graph given its eigenvalue spectrum. In this paper, we are interested in a network of technological agents whose graph is unknown, communicating by means of a consensus protocol. Recently, the use of artificial noise added to consensus signals has been proposed to reconstruct the unknown graph, although errors are possible. On the other hand, some methodologies have been devised to estimate the eigenvalue spectrum, but noise could interfere with the elaborations. We combine these two techniques in order to simplify calculations and avoid topological reconstruction errors, using only one eigenvalue. Moreover, we use an high frequency noise to reconstruct the network, thus it is easy to filter the control signals after the graph identification. Numerical simulations of several topologies show an exact and robust reconstruction of the graphs.*

## 1. INTRODUCTION

Given a network of interacting agents we describe a methodology able to reconstruct exactly the graph from time series and one eigenvalue of the graph spectrum (the inverse problem for technological networks), using recently developed signal analysis and algebraic graph theory techniques. There are many reasons to be aware of the graph topology. For example, if the graph is time-varying the knowledge of the topology is clearly of paramount relevance for the network control, especially in the mobile sensor network case. The next generation of technological networks will be equipped with topology discover methodologies.

Moreover, from a mathematical point of view, the knowledge of the adjacency or laplacian matrix of the network allows a relatively easy calculation (at least for small-medium size graphs) of fundamental parameters. To name only a few: the second largest laplacian eigenvalue (the Fiedler or algebraic eigenvalue) and the maximum eigenvalue of the adjacency matrix are two well-known parameters relevant to the connectivity, the information exchange and control. bottlenecks are easily detected; most influential nodes can be identified.

However, a fundamental obstacle to the solution of this inverse problem is the absence of one-to-one correspondence between graph and its spectrum: different graphs can have the same spectrum [1]. Although Camellas [2] has recently obtained some useful results with combinatorial optimization algorithms for the main topological parameters (diameter, average distance, degree distribution, clustering), the complete graph identification remain unsolved. As a theoretical solution seems hard to find, we propose to take advantage of specific characteristics of technological networks to reduce the complexity of the inverse problem. In particular, if the graph variability is associated to the agent mobility, a common control algorithm is the consensus protocol [4]. The consensus protocol is widely implemented because it is a distributed calculation, i.e. each agent is required to communicate only with its closest neighbours. In this paper, we consider the consensus as the main control tool of the network, though many different situation are equally possible, and following [3], we use an artificial added noise for a *basic estimation* of the topology. Moreover, we take advantage from the (approximate) knowledge of one eigenvalue of the graph spectrum, let's say the maximum eigenvalue of the laplacian matrix $\lambda_N$ *to eliminate the errors*. Some distributed procedures are already available [8, 10] to estimate the eigenvalue spectrum, as explained in the following paragraphs. Nevertheless, even the complete knowledge of the spectrum does not suffice to recover the graph, thus more information are needed. These information may be provided by the consensus time-series in the Ren's framework

In some cases these procedures are not mandatory, as the spectrum may be partially known in advance, therefore here we consider $\lambda_N$ derived by [8, 10] as a parameter of the problem affected by error and will not include it in the simulations explicitly.

Finally, it should be noted that often the consensus protocol is necessary to the technological networks, thus no further calculation encumbrance is imposed to the control system. Since also the spectrum may be calculated locally, we have two distributed elaborations, while the final calculation is a centralized one because it requires to collect data from all the nodes and send back the results to every node.

## 2. METHODS

Recently, a method to recover the laplacian matrix **L** of the a network of dynamical coupled systems has been given by Ren [3]. Starting from the general form of the *i*-th differential system:

$$\boldsymbol{x}_i' = \mathbf{F}_i(\ x_i\ )$$

$i = 1, \dots N,$ and adding couplings and noise we have:

$$\boldsymbol{x}_i' = \mathbf{F}_i(\boldsymbol{x}_i) - c\ \Sigma_j\ L_{ij}\ \mathbf{H}(\boldsymbol{x}_j) + \eta_i \tag{1}$$

$i, j = 1, \dots N$ , where $c$ is the coupling coefficient (here $c = 1$), **H** the coupling functions, $\boldsymbol{x}$ the state variables, $\eta$ the white gaussian noise with strength $\sigma^2$, $L_{ij}$ are the entries of the laplacian matrix **L** derived from the undirected graph of the systems. Vectors and matrices are in bold. The laplacian matrix is defined as:

$$\mathbf{L} = \mathbf{D} - \mathbf{A}$$

where **D** is a diagonal matrix formed by the node degrees and **A** is the adjacency matrix (1 if a link *i-j* exists, 0 otherwise) of the graph. Therefore we are considering a graph whose nodes are technological agents represented by a differential equations coupled to other nodes through communication links (the edges of the graph) according to an unknown topology.

The very interesting point in (1) is that the artificial added noise enables the solution of the inverse problem: given the time series, reconstruct the graph. We focus on the standard consensus form of (1):

$$x_i' = \Sigma_j\ a_{ij}(x_j - x_i) + \rho_i \quad , \quad j = 1, \dots N \tag{2}$$

where $a_{ij}$ are the entries of the adjacency matrix **A.** Differently from [3] we consider a high frequency (HF) noise $\rho$ instead of the white noise $\eta$; moreover, the noise strength is decreased, because in a real environment to transmit a signal requires energy.

It is known that for a connected network, the equilibrium point for (2) is globally exponentially stable and the consensus value is equal to the average of the initial values, thus solutions do exist. In compact form (2) is written:

$$\boldsymbol{x}' = -\mathbf{L}\boldsymbol{x} + \boldsymbol{\rho}$$

Expression (2) and similar are utilized to coordinate the states of the agents on a common position/velocity agreement resilient to disturbs [6, 7, 9, 14, 15, 16].

Now we give the main points of the Ren's mathematical procedure to derive the expression of the dynamical correlation matrix that solves the inverse problem. Details can be found in [3, 13]. Let us consider an undirected graph of the dynamical system (1); using small perturbations and substituting the original variables

$$x^{\circ}_i = x_i + \zeta_i \qquad i = 1, 2, \dots N$$

we obtain

$$\boldsymbol{\zeta}' = [\ \mathbf{J_F}\,(\,x^{\circ}) - \mathbf{L}^{\wedge} \otimes \mathbf{J_H}\,(\,x^{\circ})]\ \boldsymbol{\zeta} + \boldsymbol{\rho}$$

where $\mathbf{L}^{\wedge}$ is the laplacian coupling matrix, $\mathbf{J_{H,F}}$ is the Jacobian matrix of coupling functions.

The dynamical correlation among the consensus signals is called

$$\mathbf{C}^{\wedge} = <\boldsymbol{\zeta}\ \boldsymbol{\zeta}^{\mathrm{T}}>$$

After some calculations we have

$$\mathbf{A}^{\wedge}\ \mathbf{C}^{\wedge} + \mathbf{C}^{\wedge}\ \mathbf{A}^{\wedge\mathrm{T}} = <\boldsymbol{\rho}\ \boldsymbol{\xi}^{\mathrm{T}}> /\ 2 \qquad\qquad (3)$$

with $\quad \mathbf{A}^{\wedge} = -\,\mathbf{J_{F^{\wedge}}}(\,\boldsymbol{x}^{\circ}\,) + \mathbf{L}^{\wedge} \otimes \mathbf{J}_{\mathbf{H}^{\wedge}}\,(\,\boldsymbol{x}^{\circ}\,)$

(3) can be simplified as

$$\mathbf{L}^{\wedge}\ \mathbf{C}^{\wedge} + \mathbf{C}^{\wedge}\mathbf{L}^{\wedge\mathrm{T}} = \mathbf{I}\sigma^2 /\ 2$$

and therefore

$$\mathbf{C}^{\wedge} \sim \mathbf{L}^{+}\,(\sigma^2 / 2) \qquad\qquad (4)$$

where $\mathbf{C}^{\wedge}$ is the dynamical correlation matrix among the time series between node $i$ and node $j$, $^{+}$ indicates the Moore - Penrose pseudoinverse, $\sigma^2$ the noise power. Note that (3) requires the knowledge of all time series to calculate $\mathbf{C}^{\wedge}$, hence the reconstruction is centralized. Authors of

[3] find a one - to - one correspondence between the correlation matrix of time series from nodes and the laplacian matrix. This remarkable, counter intuitive finding actually allows to set a threshold for the entries of $\mathbf{C}$^: below it, the entries are considered -1, above 0, thus the laplacian matrix and consequently the adjacency matrix, is reconstructed. The threshold procedure is not immediate to implement, anyway in [3] it is claimed a very good success rate.

Albeit no physical explanation of the phenomenon is claimed simulations show good results, nevertheless, some errors are reported to remain.

Note that the Ren' s algorithm is more accurate if the average degree is large, but the energy saving needs of the signal transmission apparatus require the average degree (i.e. the number of communication links) to be kept as low as possible. As a consequence, in a real environment the $\mathbf{C}$^ estimation could be not exact.

At the same time, the consensus signals are needed also to control the network and in this respect, noise is a disturb to keep as small as possible. Therefore, bearing in mind these considerations, we suggest a node to transmit the consensus signal added with HF, low power noise and to low-pass the noisy signal received.

## 2.1 The spectral estimation

To reduce or eliminate the residual error in the graph reconstruction we need extra information. To this end, a relevant help is the knowledge at least of some eigenvalues of the laplacian spectrum. In other cases the graph is fixed and there is no need of topological variations, thus the desired spectrum is known and only a periodic verification is required, but usually the graph changes frequently and demands an on-line check.

The spectral reconstruction has been studied in [8, 10]. Franceschelli calculates a distributed Fast Fourier Transform (FFT) of signals derived from a proper distributed protocol and received at a node $i$

$$x_i' = \ z_i + \sum_j ( z_i - z_j )$$
$$z_i' = - x_i - \sum_j ( x_i - x_j )$$

with $j \in N_i$ (neighbour nodes at one hop of distance from node $i$). Thus, the state trajectory is a linear combination of sinusoids oscillating only at frequencies function of the eigenvalues of the Laplacian matrix $\lambda_j$, and the amplitude of the peaks in the spectrogram are related to the eigenvalues

$$|\mathcal{F}(x_i(t))| = 1/2 \sum_j a_{ij} \ \delta \ (f \pm (1 + \lambda_j) / 2\pi \ )$$

$$|\mathcal{F}(z_i(t))| = 1/2 \sum_j b_{ij} \ \delta \ (f \pm (1 + \lambda_j) / 2\pi \ )$$

This method has some drawbacks [11]: the multiplicities of the eigenvalues cannot be calculated and the FFT suffers from the presence of noise. Remember that independently from the Ren's procedure, communications are always polluted by several sources of noise.

On the other hand, [10] provides an estimation of the laplacian spectrum based on matrix power iteration, but this way only an approximate solution can be obtained. Thus we conclude that the methodologies of [8, 10] and similar are not completely reliable.

Finally, it is worth noting that even if an exact spectrum reconstruction were available, today is not clear if, even theoretically, it is possible to find a one-to-one relation with the adjacency matrix [12]. Alternative combinatorial optimization techniques such as the tabu search or simulated annealing are not exact and some of them would anyway require a long computation time.

## 2.2 Error reduction

Now let us consider that the largest lapalacian eigenvalue $\lambda_N$ is available by means of one of the previously described methods. It is intuitive to use it as a simple cost function, instead of the threshold procedure, to determine the non null entries of the adjacency matrix recovered by (3).

Therefore in our methodology the matrix $\mathbf{C}\hat{}$ is calculated from noisy consensus time-series and normalized. Then, starting from a convenient value, an initial adjacency matrix $\mathbf{A}$ is produced using (4), its largest laplacian eigenvalue $\lambda_N^*$ is *calculated* and subtracted to the supposed actual eigenvalue $\lambda_N$:

$$\min g(\lambda) = \mid \lambda_N - \lambda_N^* \mid$$

and when the cost function reaches the minimum

$$g(\lambda_N) = 0 \qquad\qquad (5)$$

the actual matrix $\mathbf{A}$ is reconstructed (best results have been obtained with the largest eigenvalue, although other eigenvalus may be used). In Figures 1a, b, c it is shown how the zero estimation error of the eigenvalue is reached jointly with the complete reconstruction of the adjacency matrix. On the ordinate the reconstruction error of the eigenvalue and of the adjacency matrix: when the two vertical lines coincide, the reconstruction is done.

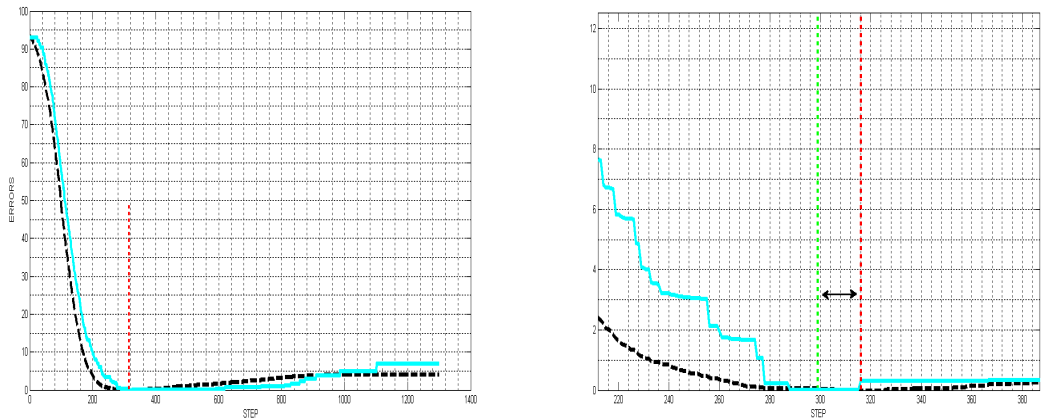

Figure 1a: Small World graph, 100 nodes (abscissa: time steps, ordinate: errors). Black dotted curve: actual error percentage of the adjacency matrix entries (not including diagonal and symmetric elements), continuous blue curve: the eigenvalue absolute error $\mid \lambda_N - \lambda_N^* \mid$. Vertical red and green (green line is not visible because is coincident with the red one) lines: exact reconstruction according to the (4). The minimum value of the continuous blue curve indicates the correct topology reconstruction, i.e. zero errors. Figure 1b: In this case two entries are wrong and the minimum indicated by the largest eigenvalue (green dotted line), is no more coincident with the actual zero reconstruction error (dotted red line), see Table 2 also.

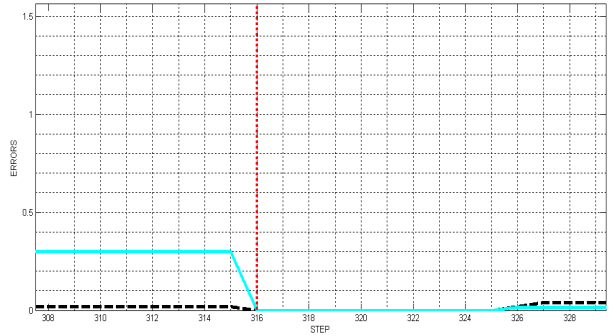

Figure 1c: Enlargement of the minimum area of Figure 3a.

If errors in the exact estimation of the maximum laplacian eigenvalue were present, the exact reconstruction as in Figure 1a is still possible for a low – moderate amount of error, although the acceptable error in the eigenvalue estimation increases quickly (see Table 2).

## 2.3 Noise addition

For the methodology to work it is necessary the addition of noise to the consensus protocol. As pointed out before, in a real environment it is already present a background of natural or artificial noise, then the noise level is further increased. This does not undermine the methodology, provided the strength of the added noise is large enough.

In order to save energy and allow the consensus signals to produce a proper control action, we add a high frequency (HF), low amplitude, zero mean, unitary variance Gaussian noise to (1). Noise strength in simulations is $\sigma^2 = 0.01$, one order magnitude smaller with respect to [3]. In Figure 2b is shown the HF noise and the signal power spectral density (psd) spectrum (frequencies are normalized). In Figure 2b, c it is shown a noisy consensus signal and as it appears after the low-pass filtering, once the signal has been received in a node. Aside the delay due to the low-pass filter, the original signal is recovered (Figure 2c).

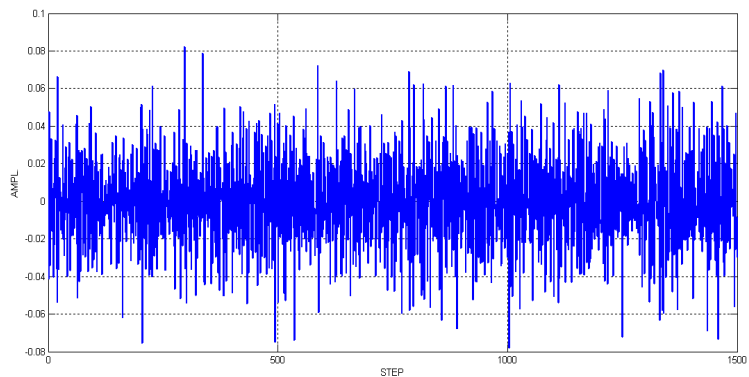

Figure 2a A consensus signal with HF noise in the time domain.

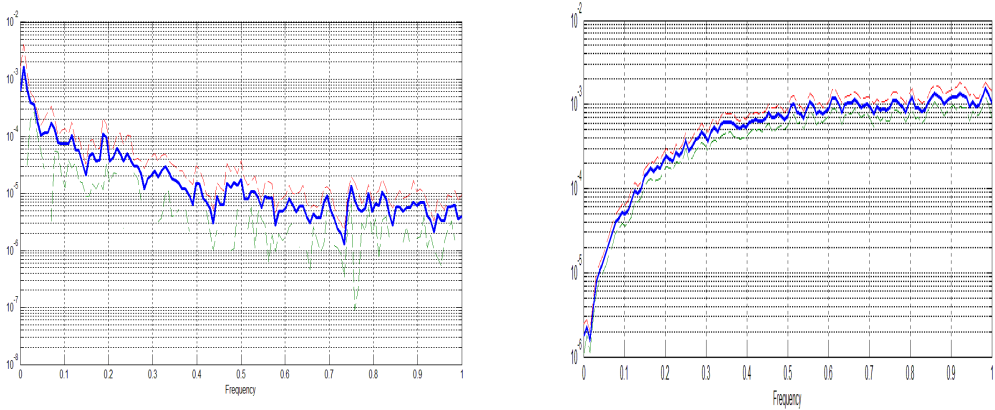

Figure 2b Power spectral density (psd) of the artificial HF noise added; left, psd of the original consensus signal with noise added.

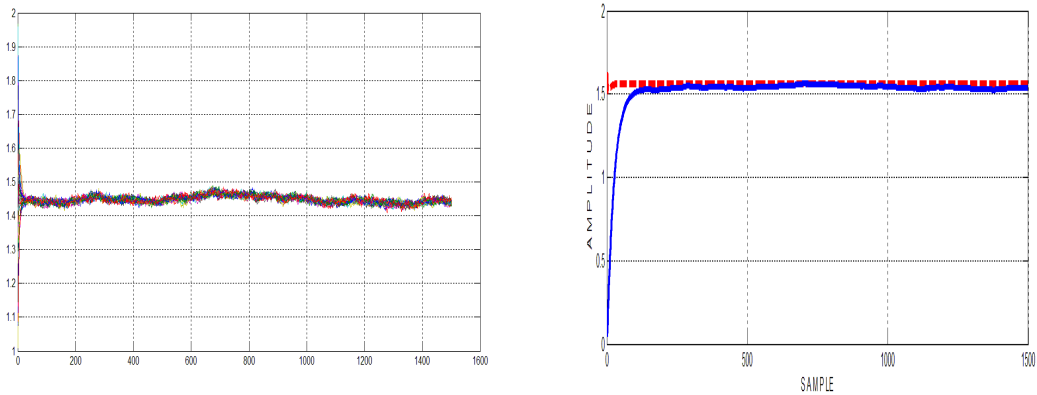

Figure 2c Noisy consensus solutions for a Small-World topology ($N = 24$, average degree 4) before the low-pass filtering. Left: consensus time series after the low-pass filtering. The red dotted curve is the original (no noise) time series.

## 2.4. Numerical simulations

Numerical simulations have been conducted to validate the methodology, results are shown in Table 1. The task is to recovery exactly all of the significant $(N^2 - N) / 2$ entries of the adjacency matrix **A** of the graph.

Four types of topologies have been considered: Erdos-Renyi (random, $p = 0.01$), Small-World (average degree: 4, $p = 0.1$), pipeline (average degree: 4), grid (average degree: 4), $N = 24$, see Figure 3. All time series have a length of 150 simulation time-steps (1500 samples) for $N = 24$; the first 30 samples have been discarded because the transitory impair the calculations. As the size (in nodes) increases, longer time series are needed. As an example, when the node size of a Small-World (SW) graph is 100, about 370 simulation steps are needed to recover the graph.

Note that an higher noise level reduces the steps, bur increases the energy dissipation. The trade-off should be analysed on an *ad hoc* basis. In Table 2 are shown the results for a large and a small SW graphs in presence of errors on the estimation of the maximum laplacian eigenvalue, obtained by the methods of [10] or [8].

The acceptable error on the maximum eigenvalue estimation (meaning that the number of mistaken entries of A is still zero) increases as $N$ increases. For example for $N = 100$, the 3.22% estimation error means that the real value $\lambda_N = 4.0375$ is altered as much as: $\lambda_N \pm 0.13$, but the reconstruction of the matrix **A** remains exact. In the simulations, the centralized elaborations are represented by the computation of the inverse correlation matrix $\mathbf{C}^\wedge$ among all the time-series received from the $N$ nodes.

For each configuration a complete reconstruction (zero errors) has been achieved, see Table 1. In particular, Small-World networks are very interesting as pointed out by [5], because of the high consensus speed and connectedness properties, but the noise addition slows down the consensus computations.

Table: 1 Simulation results

| Graph topology | Error | Nodes | Links | Integration steps |
|---|---|---|---|---|
| Erdos-Renyi | 0 | 48 | 16 | ~150 |
| Small-World | 0 | 24 | 48 | ~150 |
| Small-World | 0 | 100 | 200 | ~370 |
| Pipeline | 0 | 24 | 43 | ~150 |
| Grid | 0 | 24 | 38 | ~150 |
| Grid | 0 | 100 | 180 | ~370 |

Since we have conducted the simulation with a high level language and a non optimized code, the actual calculation time is not significant. In the real-world application a C or Java optimized programming language or a Digital Signal Processor must be used, reducing the actual calculation.

The small-world consensus scheme seems to be the fastest also for a low number of nodes. This is interesting, because although it is known [5] that when a Small-World has a number of nodes $N > 100$ the convergence is very fast, for $N = 24$, as in our case, there is no previous guarantee.

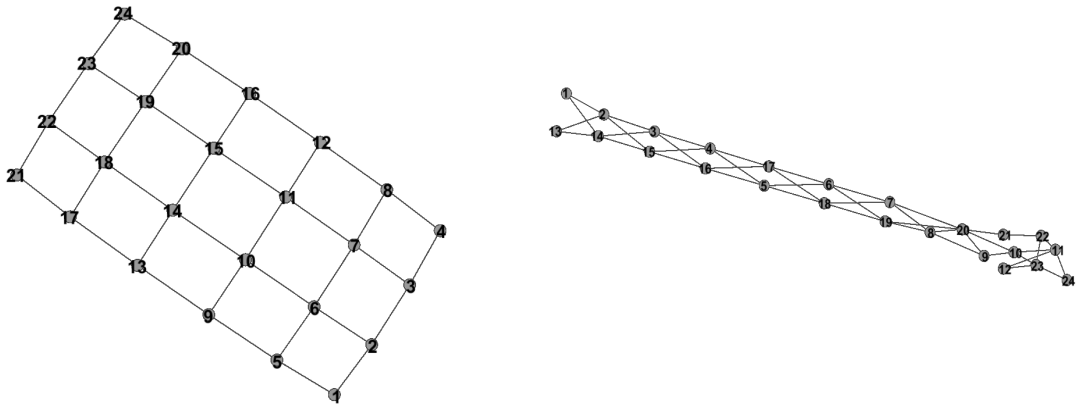

Figure 3a  Left to right graph topologies:  regular grid, pipe-line. Each node is an agent, links are wireless communication channels. The grid topology is the most regular, the SW is half-way between regularity and randomness.

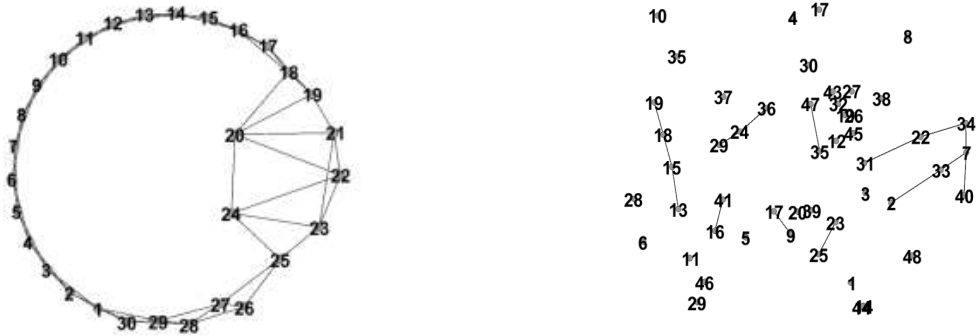

Figure 3b  Left to right graph topologies: Small-World and  random. The grid topology is the most regular, the SW is half-way between regularity and randomness. Note the some disconnected nodes of the ER random topology.

Table 2: Stability of solutions for SW topology

| Graph topology | Mistaken entries | Nodes | Overall Entries of A | Acceptable error in the $\lambda_N$ estimation |
|---|---|---|---|---|
| SW | 0 | 100 | 4950 | 3.22% |
| SW | 2 | 100 | 4950 | 7% |
| SW | 0 | 24 | 276 | 0.22% |
| SW | 2 | 24 | 276 | 15.2% |

## 2.5 Effects of delays on the reconstruction

[13] has extended the Ren's method to the quasi-uniform delay case. While the integration of stochastic delayed differential equations is not simple, from a theoretical point of view the procedure is similar to the zero delay case. Therefore, it is conceivable that the framework discussed here is suited also for the delayed case. It should be mentioned that also the threshold effects [17] may influences delays and have to be considered carefully.

## 3. CONCLUSIONS

One of the most important and unsolved problem in graph theory is the reconstruction of the topology of technological networks. In fact, position sensors are often inaccurate, unable to work properly and the graph may change suddenly. At the cost of a semi - centralized elaboration of the consensus time series, we have shown how it is possible to achieve a complete topology reconstruction. The methodology envisages the reconstruction of the graph using the noisy signals of the consensus protocol. When received, signals are correlated and the resulting correlation matrix is elaborated according to a simple relation to obtain the laplacian matrix. Since the largest eigenvalue of the laplacian matrix can be estimated independently, although not exactly, it is possible to calculate a cost function of the reconstruction. This information allows to decide the adjacency matrix with zero or minimum error. The original consensus signals necessary to the control are recovered by low-pass filtering, as noise is allocated in the relatively high frequency band.

## ACKNOWLEDGEMENTS

This work has been supported by the HARNESS Project, funded by the Italian Institute of Technology (IIT).

**Authors**

Enzo Fioriti  graduated in Automatic Control Engineering at La Sapienza Rome University. He has worked in the ICT private sector and as researcher in Complex Systems, Neural Networks Critical Infrastructures, Graph Theory at the ENEA CR Casaccia. Currently he is interested in underwater robot swarm.

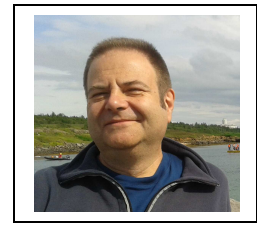

Stefano Chiesa received his bachelor's and the master degree in Computer Science from Roma 3 University, Rome, in 2006 and 2010 respectively. Currently he is working at ENEA (Italian national agency for new technologies, energy and sustainable economic development) as research fellow. His research interest include swarm robotics, formation control, computer vision, visual odometry.

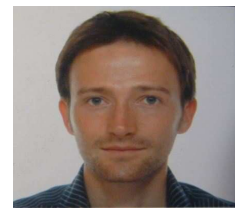

Fabio Fratichini graduated in Electronic Engineering at La Sapienza Rome University. He has worked as a researcher at ENEA (National agency for new technologies, energy and sustainable economic development) research center. He is working as researcher in robotics, sensor data fusion, Kalman filter. Currently he is interested in localization of underwater robot swarm within the HARNESS Project.

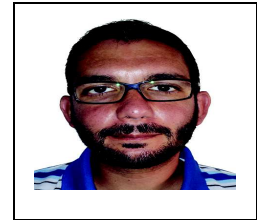